\documentclass[10pt,conference]{IEEEtran}
\IEEEoverridecommandlockouts
\usepackage{cite}
\usepackage{amsmath,amssymb,amsfonts}
\usepackage{amsthm}
\usepackage{nccmath}
\usepackage{fancyhdr}
\usepackage{svg}
\usepackage{bbm}
\usepackage{soul}
\usepackage{subfloat}
\usepackage{graphicx}
\usepackage{textcomp}
\usepackage{xcolor}
\usepackage{kotex}
\usepackage{algorithmicx,algpseudocode}
\usepackage{subfigure}
\usepackage{booktabs}
\usepackage{soul,color}
\usepackage{bbold}
\usepackage{mwe}
\usepackage{multicol}
\usepackage{multirow}
\usepackage[normalem]{ulem}
\usepackage[linesnumbered,ruled]{algorithm2e}

\newcommand{\BfPara}[1]{{\noindent\bf#1.}\xspace}

\def\BibTeX{{\rm B\kern-.05em{\sc i\kern-.025em b}\kern-.08em
    T\kern-.1667em\lower.7ex\hbox{E}\kern-.125emX}}
\begin{document}
\providecommand{\keywords}[1]{\textbf{\textit{Index terms---}} #1}
\newcommand{\tpurple}[1]{\textcolor{purple}{#1}}
\newcommand{\addrefs}{{\color{purple}{\hl{~[Add~Refs]~}}}}
\newcommand{\addtext}{{\color{blue}{\hl{~[Add~Texts]~}}}}

\title{EQuaTE: Efficient Quantum Train Engine for Dynamic Analysis via HCI-based Visual Feedback}


\author{
\IEEEauthorblockN{Soohyun Park$^{\dag}$, Won Joon Yun$^{\dag}$, Chanyoung Park$^{\dag}$, Youn Kyu Lee$^{\ast}$, Soyi Jung$^{\ddag}$, Hao Feng$^{\S}$, Joongheon Kim$^{\dag}$}
\IEEEauthorblockA{
$^{\dag}$\textit{Korea University, Seoul, Korea}
\quad\quad 
$^{\ast}$\textit{Hongik University, Seoul, Korea} \\
$^{\ddag}$\textit{Ajou University, Suwon, Korea}
\quad\quad 
$^{\S}$\textit{Intel Labs, Hillsboro, Oregon, USA}
}}

\maketitle

\begin{abstract}
This paper proposes an efficient quantum train engine (EQuaTE), a novel tool for quantum machine learning software which plots gradient variances to check whether our quantum neural network (QNN) falls into local minima (called \textit{barren plateaus} in QNN). This can be realized via dynamic analysis due to undetermined probabilistic qubit states. Furthermore, our EQuaTE is capable for HCI-based visual feedback because software engineers can recognize barren plateaus via visualization; and also modify QNN based on this information. 
\end{abstract}

\section{Introduction}\label{sec:intro}
\BfPara{Background and Motivation}
In modern software developments, various libraries and frameworks are incorporated for fast and massive software implementation, whose complexity for implementation is significantly high~\cite{tian2020evaldnn}. Therefore, we need software development tools for handling this issue.
As a result, various software development tools are utilized for intuitive understanding of the target source codes to aid software engineers, \textit{e.g.}, tensorboard, NVIDIA DIGITS~\cite{7966878}. 

Recently, the importance of software development tools for quantum machine learning (QML) is increasing. 
Nowadays, the QML software is one of the active research topics in modern noisy intermediate-scale quantum (NISQ) era~\cite{9912289,aaai23quantum,10012051,quantumnas,icml22yun}.
Along with this fast QML research innovation, the relevant software development tools are widely used, such as \textit{PennyLane}, \textit{Qiskit}, and \textit{torch-quantum}.
However, despite being widely used, they offer limited characteristics to fully utilize QML's characteristics and supremacy.
\begin{itemize}
\item \textit{Accessibility:} QML conducts training via quantum information theory. Non-professional software engineers may have challenges to fully utilize QML. It is required to design a new tool for QML software development for QML code generation and execution even if the software engineers are not with deep-dive understanding to QML. 
\item \textit{Feasibility:} QML can suffer from barren plateaus problem which is similar to the local minima in deep neural network (DNN) training. 
Therefore, it is required to design a new tool for QML software development, which should visually recognize barren plateaus; and it should also provide additional information which can be useful for software code analysis. Based on this, software engineers can re-organize the QML software codes, \textit{i.e.}, visual feedback considering human-computer interaction (HCI). Here, HCI can be realized because the human can re-organize the codes for the computer; whereas the computer can visualize the barren plateaus status to software engineers. 
\end{itemize}
There are various conventional QML software development tools to support \textit{Accessibility}. However, there are no QML software development tools for taking care of \textit{Feasibility}, to the best of our knowledge. 
Accordingly, this paper focuses on the \textit{Feasibility} in QML software development tool design. 

\BfPara{Design Concept}
The barren plateaus occur in quantum neural network (QNN) due to the operations of quantum gates as well as the superposition of quantum bits (qubits).
Here, the QNN can improve its own performance via entanglements among multiple qubits. 
The entangled quantum states are advantageous regarding information capacity, however, the multi-qubit interference increases while many quantum gates and qubits are utilized which are with high-dimensional quantum entanglement. 
This leads to the case where QML training is not possible due to barren plateaus. 
Therefore, in order to improve QML training performance, it is critical to define and implement well-designed QNN. 
Among various research contributions in order to solve barren plateaus, it has been proved that the barren plateaus is strongly associated with parameterized quantum circuit (PQC) which is used for QNN hidden layer design~\cite{mcclean2018barren}. Therefore, controlling the system size of PQC improves QML training performance. 
Software engineers will suffer from \textit{i)} quantum computing concept understanding (including the definition of barren plateaus) and \textit{ii)} identification of optimal system size of PQC for reducing entanglement interference. 

Motivated by this, we propose \textit{Efficient Quantum Train Engine (\textbf{\textsf{EQuaTE}})}, a novel QML software development tool for reporting barren plateaus occurrence. \textbf{\textsf{EQuaTE}} aids software engineers by providing an intuitive understanding of training information visualizations.



\section{The Barren Plateaus Problem in QML}
\subsection{Barren Plateaus}
The barren plateau is one of well-known interruption factors for QNN training where the gradients are dramatically and exponentially decreased due to quantum entanglement~\cite{mcclean2018barren}; and this frequently leads to local minima. 
In addition, recent research result mathematically verifies that there are many barren plateau situations during QNN training~\cite{you2021exponentially}; and even worse, it is hard to escape barren plateau situations~\cite{mcclean2018barren}.
Fortunately, a novel solution approach which can take care of the barren plateau problem in QNN has been proposed~\cite{friedrich2022avoiding,sack2022avoiding}.
However, these QNN-related discussions require high-level deep-dive quantum mechanics and quantum computing knowledge.
In order to deal with this problem in the view points of software engineers, it is essentially required to have a software development tool that can notice the barren plateau occurrences during QNN training.


\subsection{QNN vs. DNN} 
The general DNN is mathematically the sequential combination of linear transformation and non-linear activation functions. 
The architecture of QNN is fundamentally similar to the one of DNN. Here, one major difference between QNN and DNN is that the computation of QNN is conducted over Hilbert space whereas the one of DNN is conducted over real number space. Thus, QNN optimizes its own objective function in quantum states over Hilbert space.
For more details of QNN, the basic computation components of QNN are \textit{i)} quantum rotation gates for quantum state transformation and \textit{ii)} quantum controlled gates for entanglement generation with other qubits.
Based on the components, the QNN architecture consists of \textit{i)} state encoding circuit (encode classical data into quantum states, \textit{i.e.}, convert bits into qubits), \textit{ii)} parameterized quantum circuit (PQC) (compute to control a specific angle for input quantum state using quantum gates), and \textit{iii)} measurement layer (decode the quantum data into classical data for optimization). 
After this operation, the quantum state can be observed and the observation is used to minimize the loss function.
The procedures \textit{i--ii)} are linear transformations, and \textit{iii)} is a unique non-linear activation~\cite{you2021exponentially}.
It seems that the difficulties of training DNN and QNN are similar. However, the aforementioned barren plateaus make the loss gradient of QNN vanish due to qubit entanglement. 
By contrast to DNN training, in QNN, the number of barren plateaus are proportioned to the number of qubits exponentially, and the landscape of barren plateaus depends on the QNN design. In other words, QNN is hard to train, and explaining barren plateaus helps software engineers to design adequate QNNs.

\section{EQuaTE Design}
Based on the aforementioned motivation, our \textbf{\textsf{EQuaTE}} is designed under the considerations of following three aspects.
\begin{itemize}
    \item \textit{Barren Plateaus Problem Tracking:} For the design and implementation of QML applications, \textbf{\textsf{EQuaTE}} reports the barren plateaus occurrence to software engineers. Based on this, the software engineers can reduce un-stabilized QNN training performance observations, i.e., resilience and robust QNN training can be realized. 
    \item \textit{Dynamic Analysis in Software Testing:} 
    Here, it can be noted that the software testing of QML software codes cannot be done via static analysis which tests the integrity of QML software with original source codes. The reason is that the QML software codes should be evaluated while qubits inputs are fed into QNN where the qubits exist in probabilistic states before quantum measurement. Therefore, it is essential to analysis the QML source codes via dynamic analysis which tests the integrity of QML software during run-time executions of the codes.
    \item \textit{HCI-based Visualization:} 
    In order to identify this barren plateaus problem, our proposed \textbf{\textsf{EQuaTE}} tracks the dynamics of barren plateaus values while the qubit state changes. Therefore, the visualization of the barren plateaus dynamics is essentially required for easy and intuitive understanding to software engineers. Then, even if the software engineers are not familiar with quantum computing concepts, they can clearly understand the issues and current situations with QML codes. Furthermore, this visualization provides sufficient information to software engineers; and they can provide adequate feedback to \textbf{\textsf{EQuaTE}} via HCI-based visual feedback. 
\end{itemize}

\section{EQuaTE Implementation}\label{sec:parsing}

\begin{figure}
\centering
\includegraphics[width=1\linewidth]{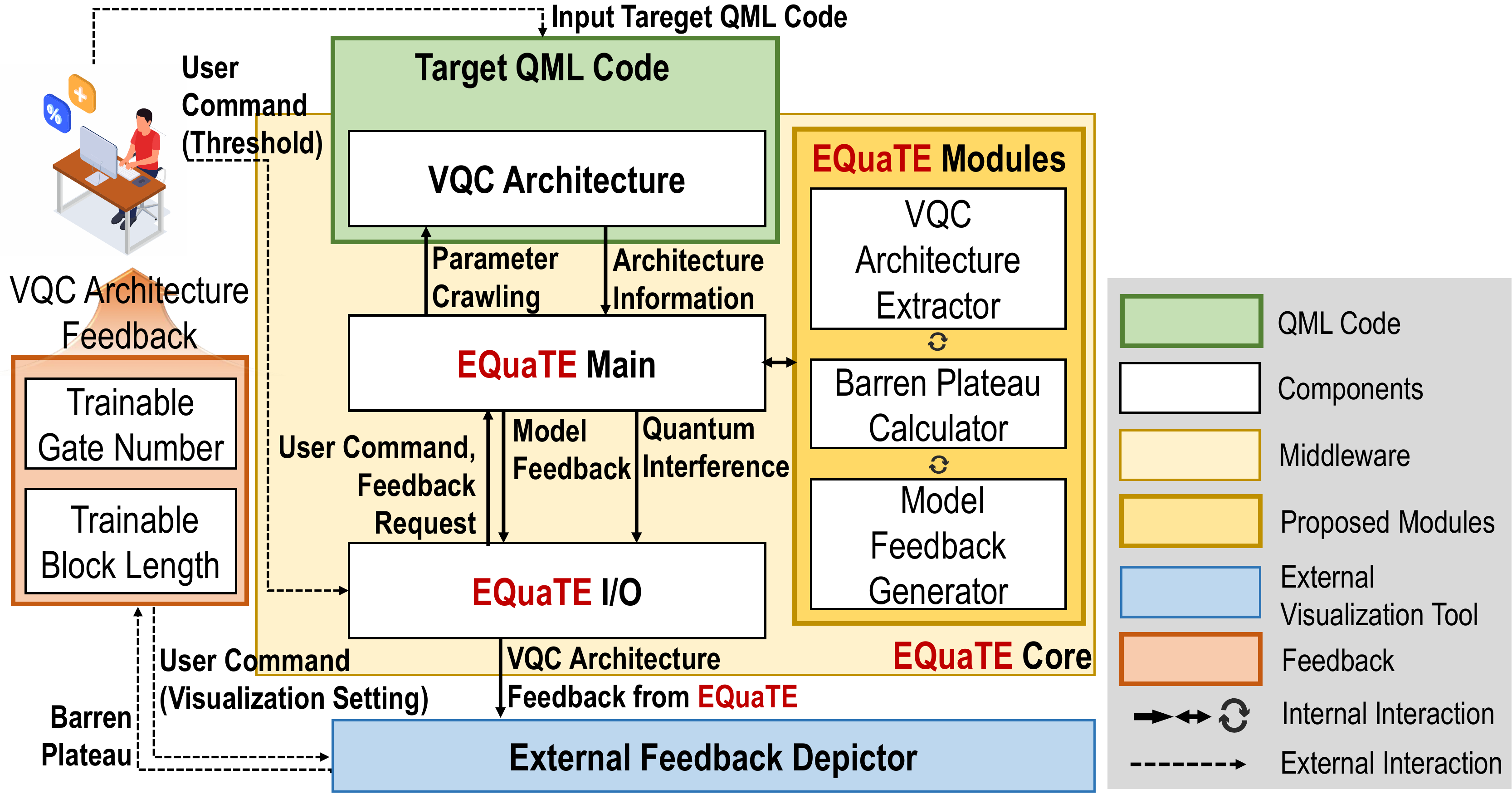}
\caption{Overall process of \textbf{\textsf{EQuaTE}} for QML application design.}
\label{fig:system_archi}
\vspace{-3mm}
\end{figure}

\subsection{Overview}

This section describes the overall architecture of our \textbf{\textsf{EQuaTE}} as illustrated in Fig.~\ref{fig:system_archi}.

\subsubsection{Key Components}
The detailed descriptions of \textbf{\textsf{EQuaTE}}'s components are as follows.
\begin{itemize}
\item \textsf{Target QML Code} corresponds to the input QNN architecture fed by a software engineer. In other words, it has a VQC architecture containing $R_X$, $R_Y$, and $R_Z$ rotation gates. Our \textbf{\textsf{EQuaTE}} can calculate the barren plateaus of the given QNN architecture with VQC.
\item \textsf{EQuaTE Main} trains the QNN model and outputs not only training results but also every gate's existence of barren plateaus calculated by the \textsf{EQuaTE Modules}. It also makes decisions about the barren plateaus occurrences and generates feedback.
\item \textsf{EQuaTE I/O} is an HCI-generating function involving the data preprocessing for transmitting the data stream to \textsf{tensorboard}. It connects bridge software engineers and computers with intuitive visualization of results.
\item \textsf{External Feedback Depictor} and \textsf{VQC Architecture Feedback} provide visualization services by serving feedback about the given QNN model in \textsf{tensorboard}. Here, \textsf{External Feedback Depictor} is a module that visualizes learning performance and barren plateaus by updating the result data every 30 seconds. In addition, \textsf{VQC Architecture Feedback} provides the software engineers with trainability information based on the barren plateaus of the currently given QNN as text.
\end{itemize}

\subsubsection{Communications}
The descriptions of \textbf{\textsf{EQuaTE}}'s communications among components are as follows.
\begin{itemize}
    \item A software engineer inputs \textit{target QML code} to our \textsf{EQuaTE Core} in order to transmit \textit{architecture information} to \textsf{EQuaTE Main} for performing QML computation.
    
    \item \textsf{EQuaTE Main} trains the given QNN model, and \textsf{EQuaTE Modules} simultaneously criticizes its trainability based on the barren plateaus at every training epoch. After the training process, it transfers \textit{model feedback} and \textit{quantum interference} to \textsf{EQuaTE I/O} to provide visualization of the results.
    
    \item \textsf{EQuaTE I/O} transmits \textit{VQC architecture feedback} (data streams about training results and model feedback) to \textsf{tensorboard} for HCI-based visualization every 30 sec.
    
    \item \textsf{External Feedback Depictor} and \textsf{VQC Architecture Feedback} visualize the transmitted data from \textsf{EQuaTE I/O} in \textsf{tensorboard}. The software engineer can realize the problems of input QNN model related to model trainability if exists. The software engineer also makes \textit{user command}, which involves a graph adjustment in \textsf{tensorboard} or the setting of \textit{threshold} of barren plateaus.
\end{itemize}

\begin{figure}
    \centering
    \includegraphics[width=1\columnwidth]{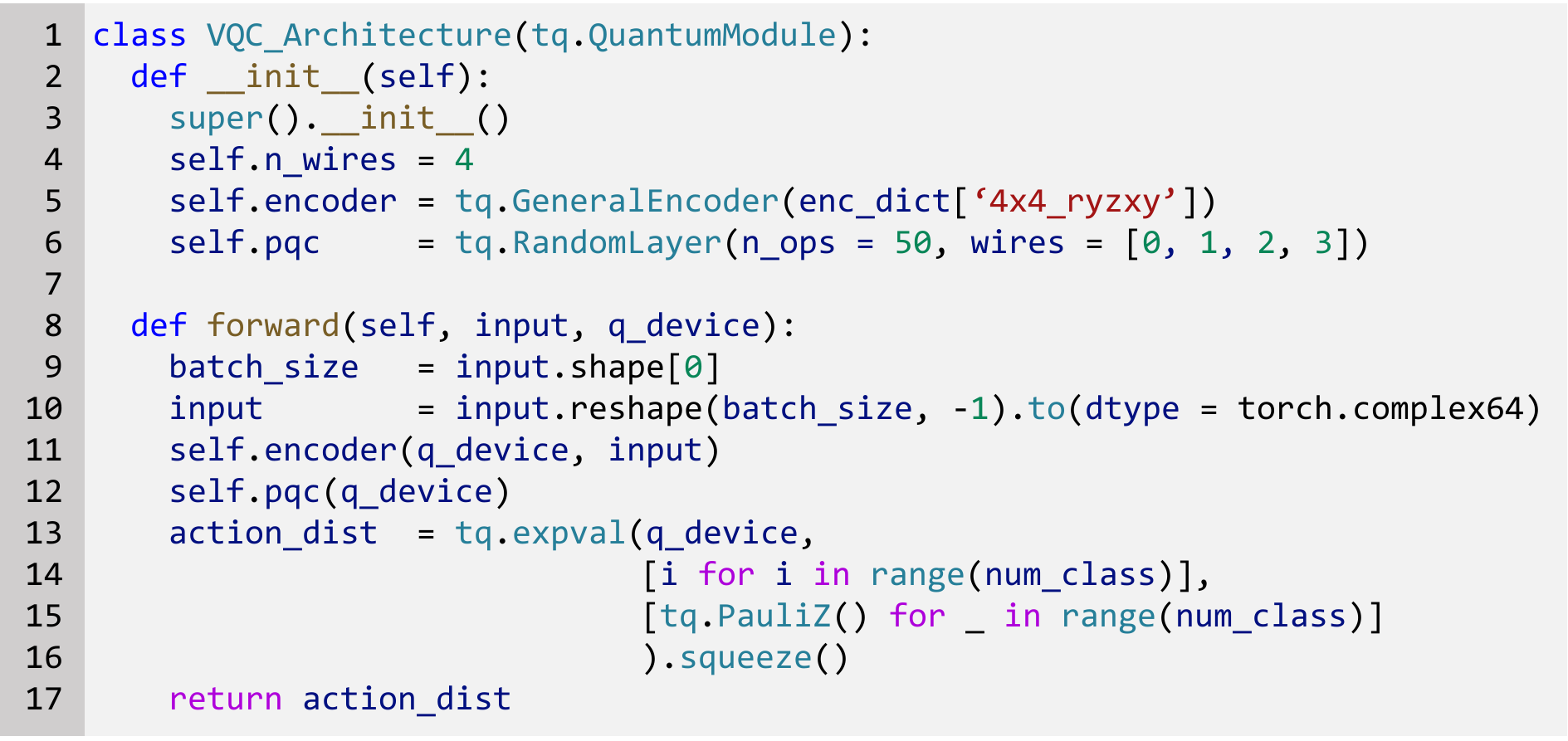}
    \caption{The sample code for \textsf{VQC Architecture}}
    \label{fig:VQC_Architecture}
\end{figure}

\begin{figure}
    \centering
    \includegraphics[width=1\columnwidth]{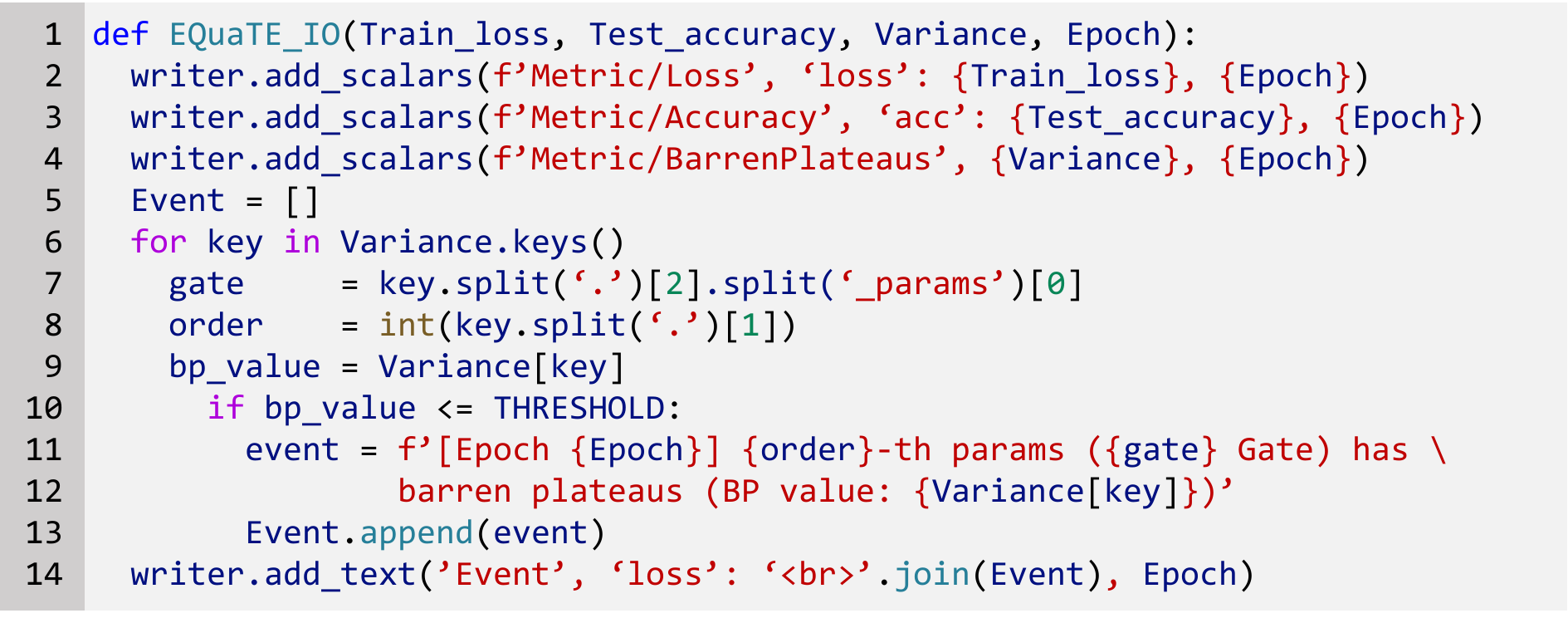}
    \caption{The sample code for \textsf{EQuaTE I/O}}
    \label{fig:EQuaTE_IO}
\end{figure}

\subsection{Implementation}
The detailed implementation of \textbf{\textsf{EQuaTE}} is as follows.

\begin{figure}
    \centering
    \subfigure[Train Loss.]{
    \includegraphics[width=0.95\columnwidth]{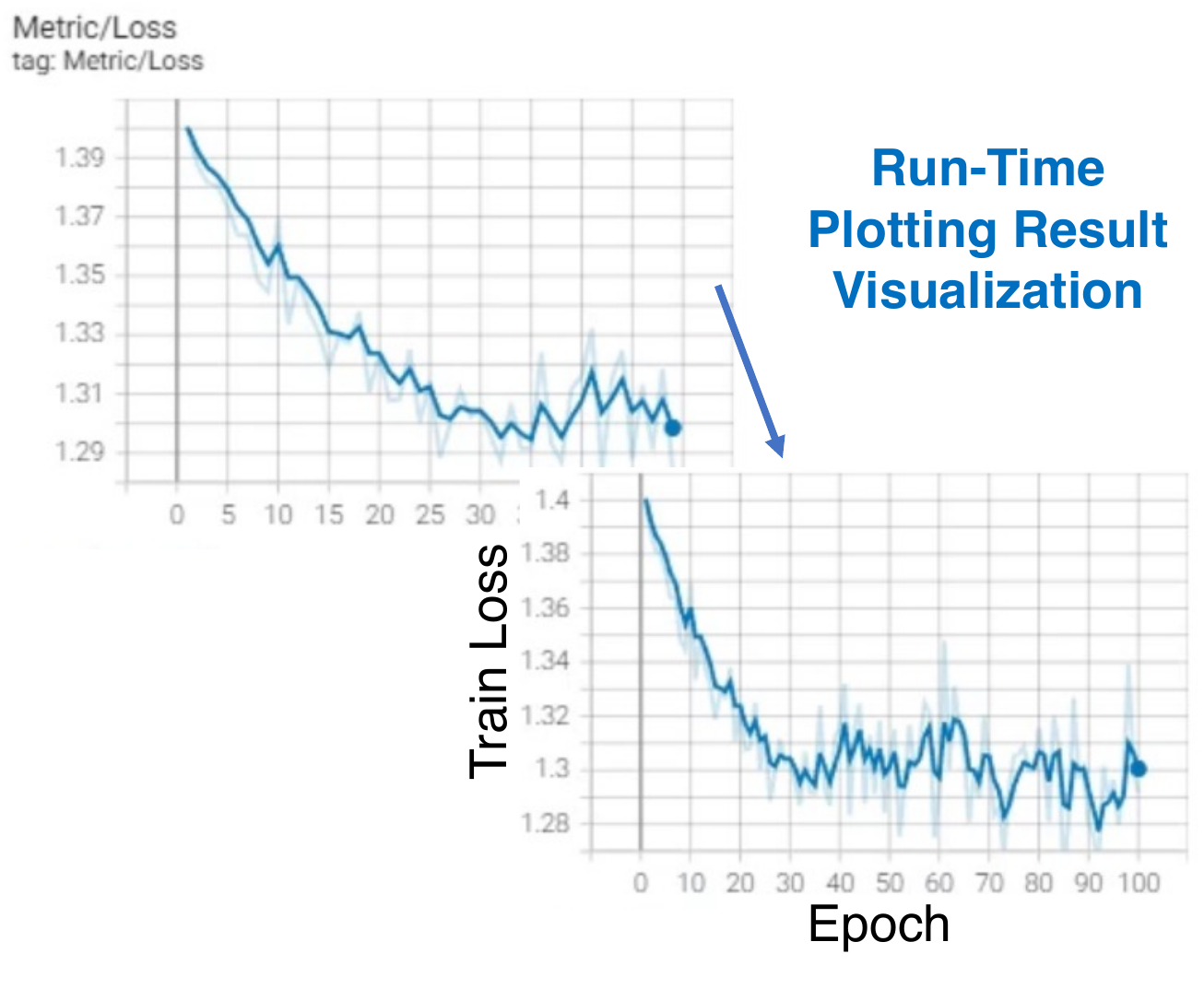}
    }
    \subfigure[Test Accuracy.]{
    \includegraphics[width=0.95\columnwidth]{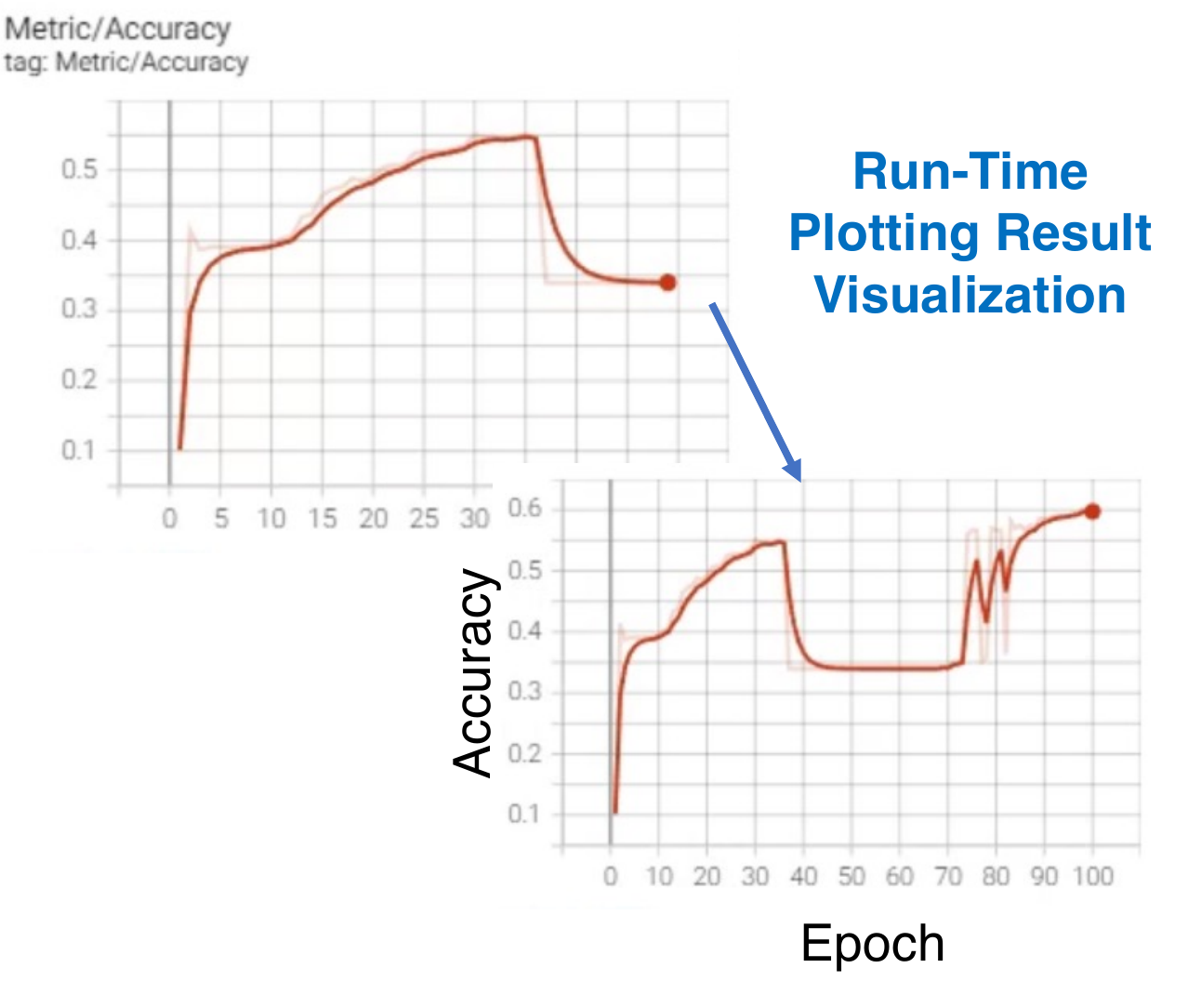}
    }
    \subfigure[Barren Plateaus.]{
    \includegraphics[width=0.95\columnwidth]{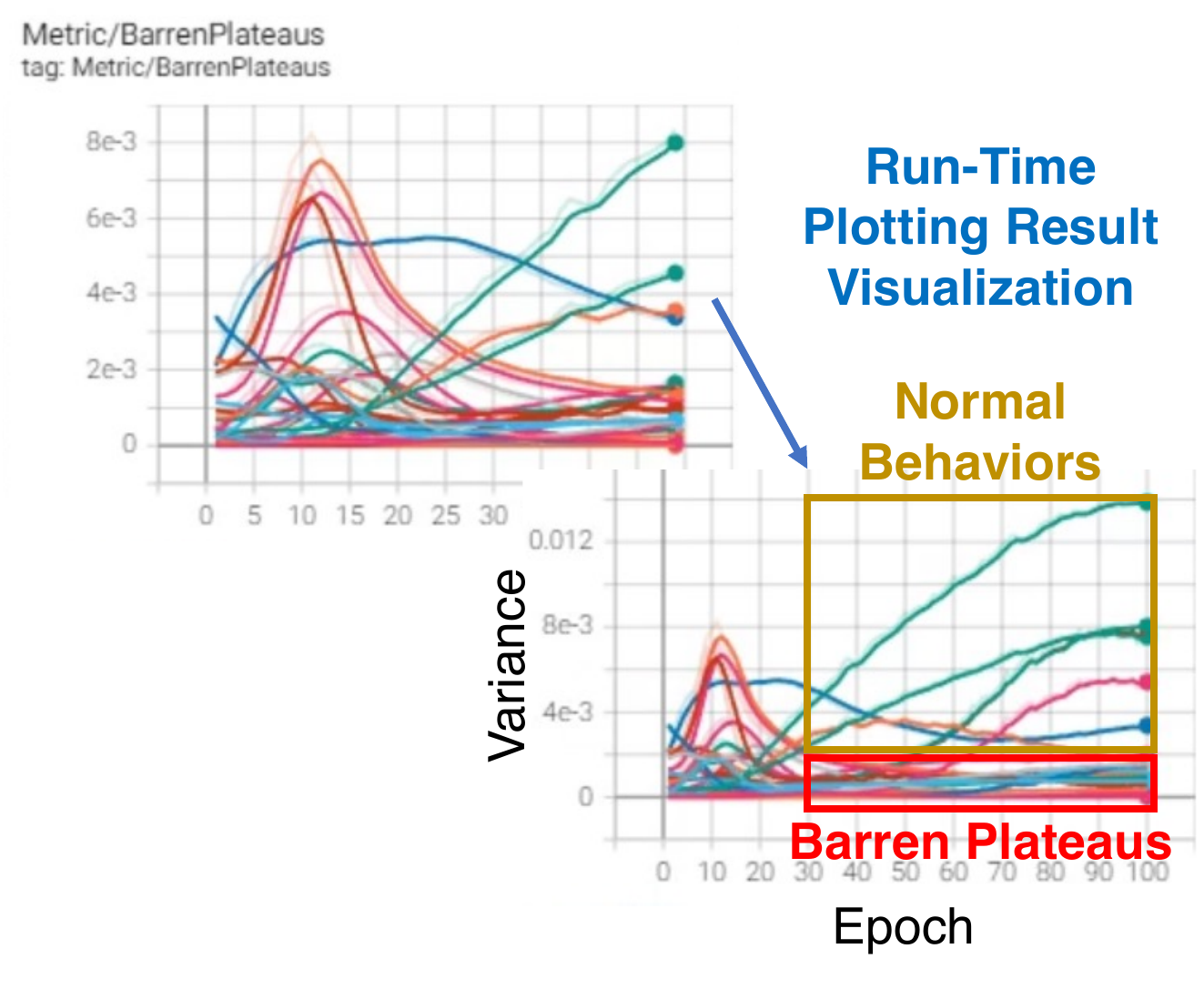}
    }
    \caption{Demonstration: training performance (captured from TensorBoard).}
    \label{fig:performance}
\end{figure}

\begin{figure}
    \centering
    \includegraphics[width=0.95\linewidth]{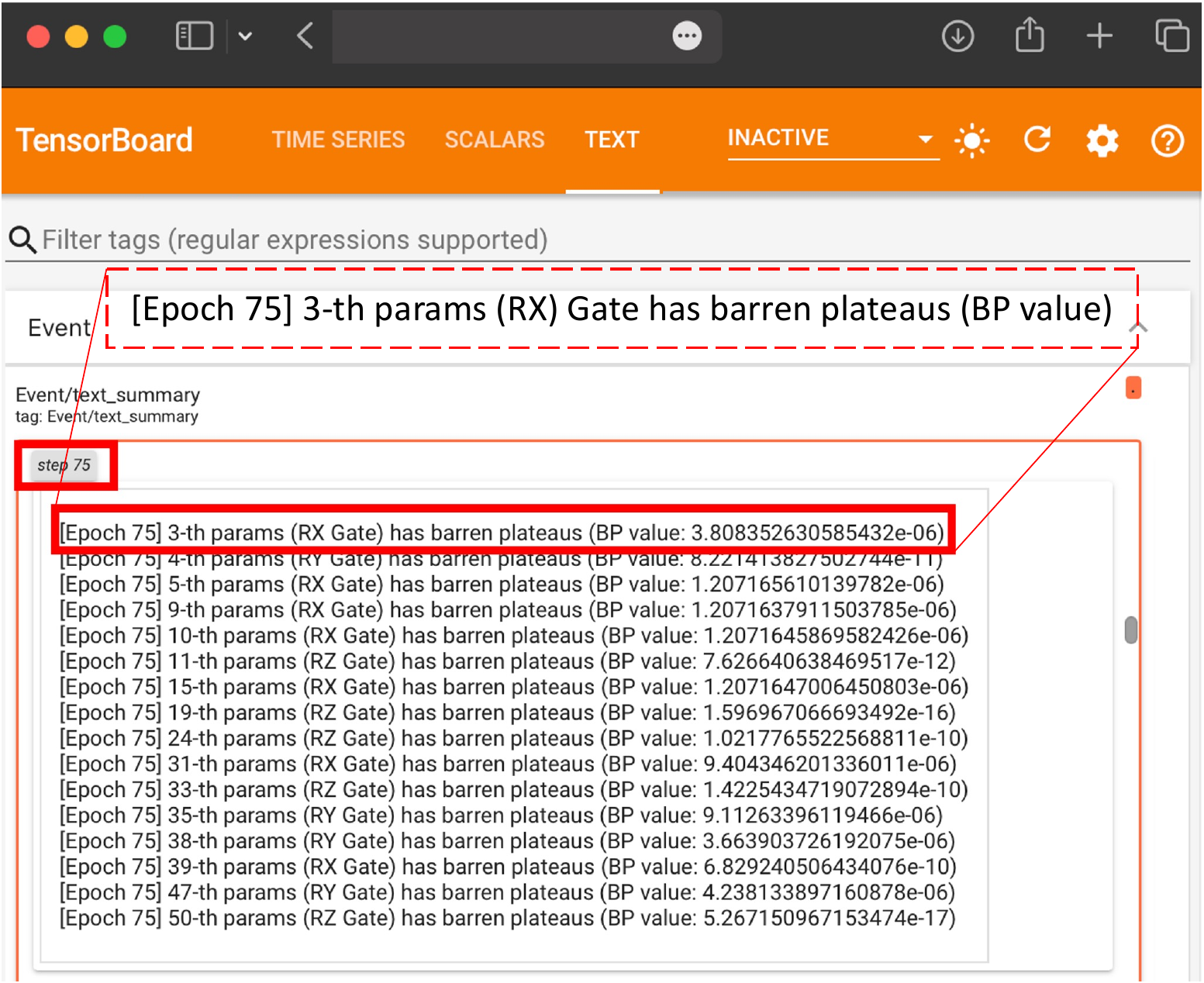}
    \caption{Barren plateaus-related QNN model feedback.}
    \label{fig:model_feedback}
    \vspace{-3mm}
\end{figure}



\begin{itemize}
    \item \textsf{VQC Architecture}: It is a quantum circuit which is the input by software engineers; and the QNN is implemented by \texttt{torch-quantum}. 
    Here, the rotation gates and controlled gates are implemented by the random layers of \texttt{torch-quantum}. The sample code for this \textsf{VQC Architecture} is in Fig.~\ref{fig:VQC_Architecture}. 
    For \textsc{(Lines 4--6)} in Fig.~\ref{fig:VQC_Architecture}, the class of \textsf{VQC Architecture} has \texttt{n\_wires} which means the number of input qubits. It also has \texttt{encoder} which transforms classical bits into qubits. In addition, \texttt{pqc} is also in this class which consists of several rotation and CNOT gates. 
    For \textsc{(Line 8--17)} in Fig.~\ref{fig:VQC_Architecture}, the \texttt{forward} function is called when the QNN object is generated. The input data for this QNN object passes \texttt{encoder} and \texttt{pqc}. 
    After that, \texttt{tq.PauliZ()} (within \texttt{tq.expval()} function) can conduct projections over $z$-axis. Lastly, the output result of measurement corresponds to QNN's decision-making. 
    \item \textsf{EQuaTE Main}: Our \textbf{\textsf{EQuaTE}} module consists of \textsf{VQC Architecture Extractor}, \textsf{Barren Plateau Calculator}, and \textsf{Model Feedback Generator}. 
    \begin{itemize}
        \item \textsf{VQC Architecture Extractor}: It calls parameter information using the \texttt{named\_parameters()} function; and the gradient calculation can be done by the \texttt{backward()} function.  
        \item \textsf{Barren Plateau Calculator}: It calculates the barren plateau values which is the variance of gradients in quantum gates. If the gradient becomes under the threshold set by a software engineer, it is determined that the computation falls into barren plateau. This is implemented by the \texttt{var()} function which is one of statistics functions. 
        \item \textsf{Model Feedback Generator}: It takes care of the event which is for the case where the barren plateau values of all quantum gates are less than pre-determined threshold; and generates texts about that where the texts include \texttt{epoch}, \texttt{parameter index}, \texttt{parameter type}, and \texttt{barren plateaus value}.       
    \end{itemize}
    \item \textsf{EQuaTE I/O}: The sample code for this \textsf{EQuaTE I/O} is in Fig.~\ref{fig:EQuaTE_IO}. It manages the inputs/outputs of \textbf{\textsf{EQuaTE}}. It sends the data for visualization to \textsf{External Feedback Depictor} (i.e., \textsf{tensorboard}). For plotting training data, the \texttt{add\_scalar()} function of \texttt{SummaryWriter} is used for the visualization of `train loss', `test accuracy', and `barren plateaus' values in order to conduct HCI-based visual feedback  \textsc{(Lines 2--4)}. For the text presentation of training data, the \texttt{add\_text()} function of \texttt{SummaryWriter} is used for 'model feedback' value presentation \textsc{(Lines 2--4, Line 14)}. 
    \item \textsf{External Feedback Depictor}: It is a component for recording the data from \textbf{\textsf{EQuaTE}} and writing to \textsf{tensorboard}. 
    Then, software engineers can identify training data via \textsf{tensorboard} and check which local parameters are in local minima. 
\end{itemize}

\subsection{Demonstration}\label{sec:demo}


The visualization results while executing our proposed \textbf{\textsf{EQuaTE}} are as shown in Fig.~\ref{fig:performance}. 
Here, 
Fig.~\ref{fig:performance}(a) and Fig.~\ref{fig:performance}(b) stand for the training loss and test accuracy during QNN task (i.e., classification) training.  
Fig.~\ref{fig:performance}(c) stands for the variance values in $R_X$, $R_Y$, and $R_Z$ rotation gates in VQC in each epoch during QNN training. 
If the variance values are less than pre-determined threshold, it means the corresponding gates are in barren plateaus. Furthermore, Fig.~\ref{fig:performance}(c) also shows the results with the form of run-time plotting result visualization, thus, intuitive understanding can be available for software engineers even for the case where they are not familiar with QML.
Lastly, our \textbf{\textsf{EQuaTE}} presents which gates are in barren plateaus with text format, as shown in Fig.~\ref{fig:model_feedback}.

In summary, software engineers can observe QNN training performance via our proposed \textbf{\textsf{EQuaTE}}. If the performance of current QNN training is not good enough, the software engineers can identify the problematic quantum rotation gates in quantum circuits using 'Model Feedback' in \textbf{\textsf{EQuaTE}}. Based on this, the software engineers can adjust the parameters of the problematic gates for improving QNN training performance. 

\BfPara{Demonstration Video and Code} 
The video demonstration and code for our \textbf{\textsf{EQuaTE}} are in~\cite{demovideolink,githublink}. The demo shows the entire process of implementing our \textbf{\textsf{EQuaTE}}. We also demonstrate various visualization services in \textsf{tensorboard}.

\section{Data Availability}
Our software implementation can be downloaded~\cite{demovideolink,githublink}.

\section{Concluding Remarks}\label{sec:conclusion}
This paper proposes a novel QML software tool which visually plots gradient variances (for checking whether barren plateaus occurs) via dynamic analysis, called \textbf{\textsf{EQuaTE}}. Furthermore, our \textbf{\textsf{EQuaTE}} is capable for visual feedback because software engineers recognize barren plateaus using tensorboard; and thus modify QNN structures based on that. 

\section*{Acknowledgement}
This work was supported by the National Research Foundation of Korea (2022R1A2C200486). Youn Kyu Lee, Soyi Jung, and Joongheon Kim are corresponding authors (e-mails: younkyul@hongik.ac.kr, sjung@ajou.ac.kr, joongheon@korea.ac.kr).

\bibliographystyle{IEEEtran}
\bibliography{refs,ref_quantum}
\end{document}